\documentclass[12pt, draftclsnofoot, onecolumn]{IEEEtran}
\usepackage{mathtools}
\usepackage{amsmath,color,amsthm}
\usepackage{breqn}
\usepackage{bm}
\usepackage{cite}
\usepackage{algorithm,algpseudocode}
\usepackage{caption}
\usepackage{lipsum}
\usepackage{cuted}
\usepackage{enumerate}
\usepackage{booktabs} 
\usepackage{graphicx}
\usepackage{float}
\usepackage{mathrsfs}

\usepackage{algpseudocode}
\usepackage{algorithm}
\algnewcommand\algorithmicforeach{\textbf{for each}}
\algdef{S}[FOR]{ForEach}[1]{\algorithmicforeach\ #1\ \algorithmicdo}
\usepackage{eqparbox}
\newdimen{\algindent}
\algnewcommand\LeftComment[2]{%
\hspace{#1\algindent}$\triangleright$ \eqparbox{COMMENT}{#2} \hfill %
}
\algnewcommand\LeftCommentNoTriangle[2]{%
\hspace{#1\algindent} \eqparbox{COMMENT}{#2} \hfill %
}

\usepackage{amsmath,amsthm,amssymb}
\DeclareMathOperator*{\argmax}{argmax}

\usepackage{wasysym} 

\DeclarePairedDelimiterX\Basics[1](){ #1}

\begin{document}

	\title{Simultaneous Energy Harvesting and Information Transmission in a MIMO Full-Duplex System: A Machine Learning-Based Design}
	\author{Yasser Al-Eryani, Mohamed Akrout, and Ekram Hossain \thanks{Y. Al-Eryani and E. Hossain are with the Department of Electrical and Computer Engineering at the University of Manitoba, Canada (emails: aleryany@myumanitoba.ca, Ekram.Hossain@umanitoba.ca). Mohamed Akrout is affiliated with the Department of Computer Science at the University of Toronto, Canada (email: makrout@cs.toronto.edu). 
The work was supported by a Discovery Grant form the Natural Sciences and Engineering Research Council of Canada (NSERC).}}

	\maketitle{}
	\begin{abstract}
	 We propose a multiple-input multiple-output (MIMO)-based full-duplex (FD) scheme that enables  wireless devices to simultaneously transmit information and harvest energy using the same time-frequency resources. In this scheme, for a MIMO point-to-point set up, the energy transmitting device simultaneously receives information from the energy harvesting device. Furthermore, the self-interference (SI) at the energy harvesting device caused by the FD mode of operation is utilized as a desired power signal to be harvested by the device. For implementation-friendly antenna selection and MIMO precoding at  both the devices, we propose two methods: (i) a sub-optimal method based on  relaxation, and (ii) a hybrid deep reinforcement learning (DRL)-based method, specifically, a deep deterministic policy gradient (DDPG)-deep double Q-network  (DDQN) method. Finally, we study the performance of the proposed system under the two implementation methods and compare it with that of the conventional time switching-based simultaneous wireless information and power transfer (SWIPT) method.
	Findings show that the proposed system gives a significant improvement in spectral efficiency compared to the time switching-based SWIPT. In particular, the DRL-based method provides the highest spectral efficiency. Furthermore, numerical results show that, for the considered system set up, the number of antennas in each device should exceed three to mitigate self-interference to an acceptable level.     
	\end{abstract}
\section{Introduction}
\subsection{Background and Related Work}
Spectrally-efficient communication methods will be crucial to support extremely massive machine type communications (EmMTC) in beyond 5G (B5G)/6G cellular networks \cite{DOMA,IoT,GCoMP,5G}.  
Multiple-input multiple-output  (MIMO)-enabled point-to-point (P2P) communications that uses either the cellular spectrum or the unlicensed spectrum is viewed as an enabling technology to increase the transmission rate \cite{D2DRef1,MIMORef1,D2D_MIMO}. 
This is achieved by MIMO beamforming and spectrum reuse of the P2P system (a special case of which is the MIMO device-to-device [D2D] system) \cite{P2P_1, D2D_MIMO}.
However, adding more antennas at each terminal device results in more hardware and processing requirements, and hence, more power consumption \cite{D2D_MIMO_Energy_Efficiency}. 
The energy consumption issue becomes more critical when terminal devices does not have a permanent power supplies and are located in remotely (e.g. distributed wireless sensor nodes, health-monitoring sensors, sensors in surveillance systems). This issue can be tackled by equipping remote wireless devices with an energy harvesting units (EH) that is able to scavenge bearing electromagnetic (EM) radio frequency (RF) signals, convert it into an energy \cite{Paradiso2005,Yasser, EHRef1, 7572107}. 
The main idea behind wireless power transfer is that the received EM RF signal is transformed from the receiving antenna to an analog-to-digital (A2D) converting unit (rectifier followed by frequency down converter). The A2D unit either stores the harvested energy at a battery or uses it to process and/or transmit an information-bearing signal (IS). EH can be conducted throughout two main paradigms namely; time switching and power splitting \cite{Paradiso2005}.
In the time-switching paradigm, the transmission time slot is splitted into two time durations: one for EH and the other for information transfer (IT). This scheme reduces the spectral efficiency by a factor equals to the ratio between EH and IT durations. In contrast, with power splitting, a part of the received signal is converted into an energy while the other part is decoded as a desired message signal \cite{6567869,Zhang2013}.    

To further enhance the spectrum efficiency, recent developments on P2P communications have adopted the use of full-duplex (FD) communication setup \cite{Paradiso2005,FD_D2D}. 
In P2P-FD communications, two RF communicating devices transmit and receive simultaneously using the same frequency band \cite{FD_1}. 
Theoretically, FD radio can double the spectral efficiency of a communication system due to concurrent utilization of same spectrum by two devices \cite{FDD}. 
One major drawback of using FD radio is the high self-interference (SI), which, if not tackled properly, may cause a total blockage of the service \cite{FDRef1}. 
Many techniques were proposed to mitigate  the effect of SI in an FD system \cite{FDRef1,FDRef2}. MIMO transmissions, however, can compensate for the SI caused by the FD communications. Theoretically, a MIMO system can provide a $K$-fold increase in transmission capacity, where $K$ is the rank of the MIMO channel matrix, when the MIMO channel matrix is a full-rank matrix \cite{Huang:2011:MCC:2161741}.

In the literature, only a few works have studied the MIMO systems in an FD setup. In \cite{MIMO_D2D_FD_1}, a detailed implementation with application possibilities of a compact MIMO-P2P system under the FD mode was investigated. It was found that, with the FD operation, a user is able to achieve a throughput gain if the user is equipped with an efficient self-interference mitigation mechanism. In \cite{MIMO_D2D_FD_2}, FD-P2P devices were deployed in a cellular multi-user MIMO network where every pair of P2P devices utilizes channel diversity by leveraging cooperative transmission mode. It was found that cooperation among MIMO-P2P systems can compensate for the residual-self interference  that occurs due to the FD mode of operation. Furthermore, in \cite{Huberman2015}, a multi-user MIMO-FD precoding scheme over orthogonal frequency division multiplexing (OFDM) scheme was designed where it was found that using MIMO in an FD-P2P system enables joint signal precoding and SI cancellation with satisfactory performance. Additionally, in \cite{MIMO_EH_P2P}, the authors proposed channel estimation methods for a time switching-based MIMO-P2P system for wireless energy transfer. 

\subsection{Motivation and Contributions}
In the existing literature, a very limited amount of research has focused on the design of MIMO-FD systems for wireless IT. To the best of our knowledge, no research has focused on designing MIMO-P2P-FD systems for simultaneous wireless information and energy transfer. Also, as has already been mentioned, the traditional time switching-based wireless information and energy transfer methods result in reduced spectrum efficiency due to the dedication of a portion of transmission time slot for energy harvesting. With power splitting-based  simultaneous wireless information and energy transfer, the probability of outage increases for the data signals due to the reduced received signal to interference-plus-noise ratio (SINR). 
Furthermore, most of the precoding algorithms proposed for MIMO-P2P-FD systems are based on sequential precoding that requires significantly high hardware capabilities and takes long time to converge. 
This high processing requirements make the MIMO-P2P-FD systems impractical for beyond 5G/6G networks with a massive number of wireless devices.

Motivated by the aforementioned problems, we propose a new MIMO-P2P-FD scheme for simultaneous wireless information and energy transfer, which is different from the conventional simultaneous EH and IT schemes (e.g. time switching and power splitting). In the proposed scheme, a MIMO-FD system is designed to simultaneously perform EH and IT between two devices, where one device transmits an energy signal and receives an information signal, while the other device receives the energy signal and transmits the information signal using the same 
time-frequency resource. This is achieved by proper antenna allocations and precoding (i.e. power allocation) at both the devices (i.e. to transmit energy and receive information at one device, and to receive energy and transmit information at the other device). The objective of antenna allocation and precoding  is to optimize the spectrum efficiency of IT in the MIMO-P2P-FD system under the assumption that the device transmitting information signals depends only on the  energy harvested from the other device. 
Furthermore, for real-time implementation of both dynamic antenna allocation and precoding, we develop a deep reinforcement learning (DRL)-based method, namely, the hybrid Deep Deterministic Policy Gradient (DDPG)-deep double Q-network(DDQN) method. 

The major contributions on this paper can be summarized as follows:
\begin{itemize}

 \item We propose an antenna allocation-based EH scheme for MIMO-P2P systems that simultaneously performs EH in one direction and IT in the other direction (FD method). 
 
 \item We formulate the general problem to jointly optimize the antenna allocations and the precoding matrices at both the devices such that the weighted sum of the IT rate and the amount of harvested energy is maximized. 
\item For the proposed system, we develop a sub-optimal antenna allocation scheme that allocates antennas between EH and IT and forms two MIMO subsystems (one for IT on one side and second for energy transmission in the other side) followed by a sub-optimal MIMO precoding scheme for each MIMO subsystem (i.e. power allocation scheme).
\item For practical online implementation, we propose and design a novel hybrid DRL method, which is based on DDPG-DDQN model.
\item Finally, we study the performance of the proposed system under the two implementation methods and compare it with the conventional time switching-based simultaneous wireless information and energy transfer method.  
\end{itemize}
 
The rest of this paper is organized as follows: Section II presents the system model and the formulation of the optimization problem for antenna allocation and precoding. A sub-optimal method for antenna allocation  precoding  is presented in Section III. Section IV  proposes the hybrid DDPG-DDQN model to solve the antenna allocation and precoding problem. In Section V, simulation results for the proposed system are discussed before the paper is concluded in Section VI. 

\textit{Notations}: For a square matrix $\mathbf{S}$, $\text{Tr}\left(\mathbf{S}\right)$, $|\mathbf{S}|$, $\mathbf{S}^{-1}$, $\mathbf{S}^{\frac{1}{2}}$, $\mathbf{S}^{\dagger}$ denote its trace, determinant, inverse,  square-root and conjugate transpose, respectively. Matrices are denoted by bold-face capital letters, vectors are denoted by bold-face small letters, and the elements are denoted by small letters. $\mathbb{C}^{N \times M}$ denotes the space of $N \times M$ matrices with complex entries. $||\mathbf{z}||$ is the Euclidean norm of a complex vector $\mathbf{z}$, and $|z|$ is the absolute value of a complex scalar $z$. log(.) function has base-2 by default. 
\section{System Model and Problem Formulation}
\subsection{MIMO-P2P-FD System Model for Energy Harvesting and Information Transfer}
\begin{figure}[!htb]
	\centering
	\includegraphics[height=4.8 cm, width=8.8 cm]{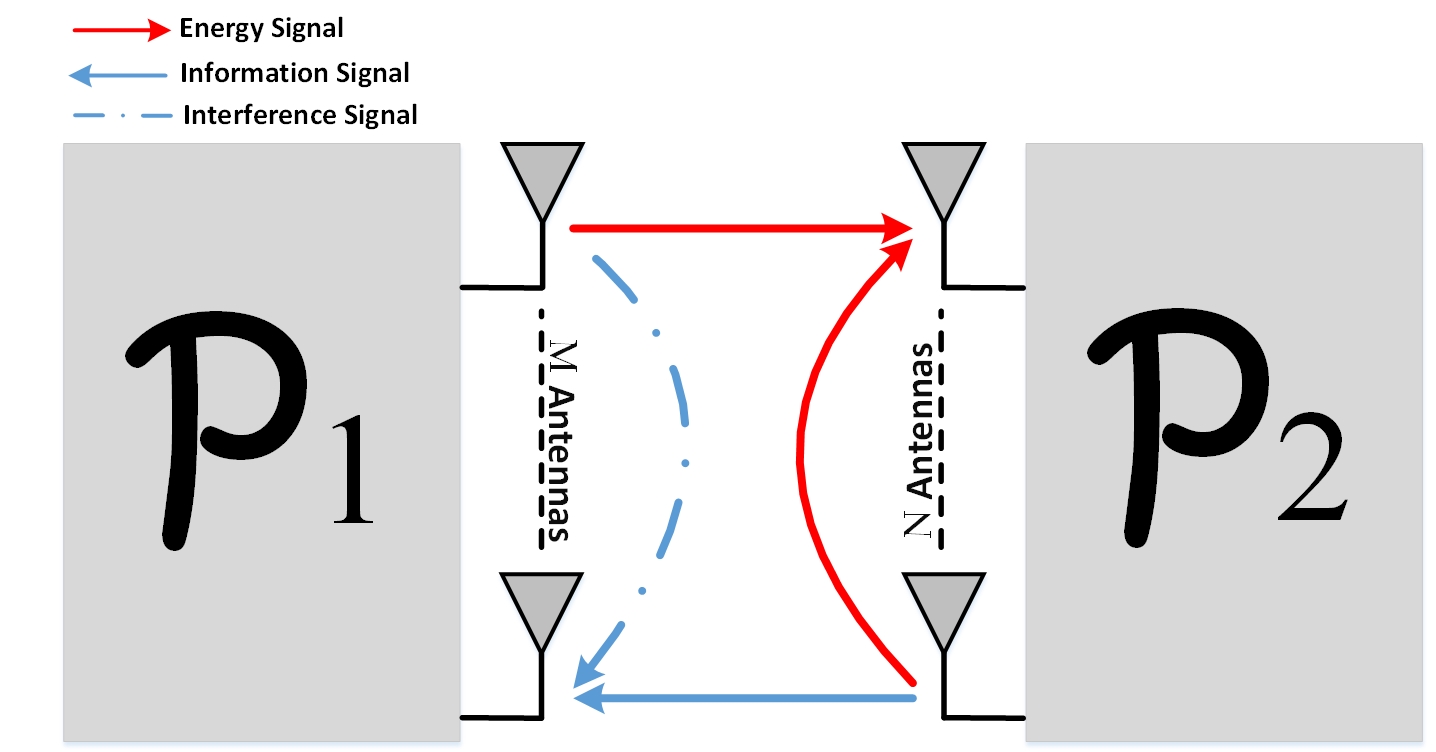}
	\caption{A MIMO-P2P-FD system for simultaneous energy harvesting and information transfer.}\label{SystemModel}
\end{figure}

Consider a pair of  devices, $\mathcal{P}_1$ and $\mathcal{P}_2$ that are simultaneously communicating with each other through a single radio frequency (RF) band as shown in Fig.~\ref{SystemModel}. 
$\mathcal{P}_1$ and $\mathcal{P}_2$ are equipped with $M$ and $N$ antennas, respectively. 
It is assumed that $\mathcal{P}_2$ has no permanent power supply and instead, it is equipped with an EH unit that harvests energy from an energy-bearing signal transmitted from $\mathcal{P}_1$, and in return, it transmits an information signal to $\mathcal{P}_1$.
On the other hand, 
$\mathcal{P}_1$ is assumed to simultaneously send an energy signal to $\mathcal{P}_2$ and receive an information signal from $\mathcal{P}_2$ using the same frequency band, i.e. it operates in FD mode.
This is achieved by properly splitting the $M\times N$ MIMO system (with channel matrix $\mathbf{H}\in \mathbb C^{N\times M}$) into two MIMO subsystems, namely, a $M_h\times N_h$ power transfer MIMO system (with channel matrix $\mathbf{{H}}_h\in \mathbb{C}^{N_h \times M_h}$) for the $\mathcal{P}_1\rightarrow \mathcal{P}_2$ link and an $M_I\times N_I$ IT MIMO system (with channel matrix $\mathbf{H}_I \in \mathbb{C}^{M_I \times N_I}$) for the $\mathcal{P}_2\rightarrow \mathcal{P}_1$ link.

The channel matrix in the direct two-way $\mathcal{P}_1\leftrightarrow\mathcal{P}_2$ link $\mathbf{H}\in \mathbb C^{N\times M}$ is assumed to be transpose symmetric and the channel gains follow a Rayleigh distribution. The channel is slowly varying such that the channel gains remain unchanged during two consecutive time slots. Accordingly, both devices $\mathcal{P}_1$ and $\mathcal{P}_2$ will suffer from SI at their receiving antennas with corresponding channel matrices $\mathbf{\mathcal{H}}_1\in \mathbb{C}^{M_I\times M_h}$ and $\mathbf{\mathcal{H}}_2\in \mathbb{C}^{N_h\times N_I}$, respectively.
Additionally, $\mathcal{P}_2$ uses its SI signal as a desired charging component for the next time slot. However, the maximum amount of harvested power at  $\mathcal{P}_2$ is limited by the maximum allowable transmission power at $\mathcal{P}_1$.

At every time slot, the MIMO subsystems are updated  with different anttenna elements  based on current channel state information (CSI) matrices $\mathbf{\mathcal{H}}_1\in \mathbb{C}^{M_I\times M_h}$ and $\mathbf{\mathcal{H}}_2\in \mathbb{C}^{N_h\times N_I}$, respectively. The elements of $\mathbf{\mathcal{H}}_1$ and $\mathbf{\mathcal{H}}_2$ are assumed to follow Rician fading distribution due to the existence of a line-of-site (LoS) component in such short unobstructed links. 
Initially, $\mathcal{P}_2$ is assumed to be fully charged with an amount of power equal to the maximum power budget of $\mathcal{P}_1$. Subsequently, at any arbitrary time slot $t$, $\mathcal{P}_2$ uses the energy harvested at previous time slot $t-1$ to send an information signal to $\mathcal{P}_1$ in the same time slot $t$. 


Let us denote by $\bm{\mathcal{C}}=\{\mathcal{C}_1 \dots, \mathcal{C}_{\delta}, \dots, \mathcal{C}_{\Delta} \}$ the set of all possible MIMO subsystem configurations such that each of the MIMO  subsystems contains at least one transmitting-receiving antenna pairs to perform energy transmission/reception (i.e. for EH) and information reception/transmission (i.e. for IT). 
As an example, with $M = 8$ and $N = 4$, one possible set is
\[
\mathcal{C}_{\delta}=\{ \overbrace{ \{ \underbrace{ \{ m_3, m_4, m_7 \}}_{M_h = 3},  \underbrace{\{n_1 \}}_{N_h = 1}\}}^{\text{EH MIMO Subsystem} }, \overbrace{\{\underbrace{\{m_1,m_2,m_5,m_6,m_8\}}_{M_I =  5},\underbrace{\{ n_2,n_3,n_4 \}}_{N_I = 3}\}}^{\text{IT MIMO Subsystem}}\}.
\]
where $m_i, i = 1,\dots,8$ and $n_j, j=1, \dots, 4$ are the indices of the antennas at $\mathcal{P}_1$ and $\mathcal{P}_2$, respectively. Note that a trade-off exists between the amount of energy harvested by $\mathcal{P}_2$ and the amount of
data received by $\mathcal{P}_2$. If more antennas are allocated for the EH MIMO subsystem,
fewer antennas will be available for IT MIMO subsystem, and vice versa. However, more power available at $\mathcal{P}_2$ (harvested from $\mathcal{P}_1$) will result in a higher information reception rate at $\mathcal{P}_1$. Accordingly, the trade-off between the number of antennas allocated for EH and IT MIMO subsystems must be optimized such that the spectrum efficiency for IT is maximized in the system.

\subsection{General Problem Formulation}
The optimal set of MIMO subsystems configurations and their corresponding precoding matrices must be the ones that produce the best system performance. Note that a high information rate at $\mathcal{P}_1$ implies that the energy harvested at $\mathcal{P}_2$ during previous time slot is high as well. Specifically, at any arbitrary time slot $t$, every set of MIMO subsystem configurations $\mathcal{C}_{\delta}, \forall \delta=1, \dots, \Delta$ will have the corresponding optimal precoding  matrices $\mathbf{W}_1^*\in \mathbb{C}^{M_h\times M_h}$ and $\mathbf{W}^*_2\in \mathbb{C}^{N_I\times N_I}$ at $\mathcal{P}_1$ and $\mathcal{P}_2$, respectively. Such an optimization process may be performed at device $\mathcal{P}_1$ at which a permanent power supply is assumed to be available. 
For the system to be implemented, $\mathcal{P}_1$ and $\mathcal{P}_2$ must be equipped with at least two antennas each. Furthermore, every antenna at $\mathcal{P}_1$ and $\mathcal{P}_2$ should either transmit/receive energy/information once at a time.

At any arbitrary time slot, the received signals at $\mathcal{P}_1$ and $\mathcal{P}_2$ under the $\mathcal{C}_{\delta}$-th MIMO subsystem configurations, are given, respectively, by
\begin{equation}
    \mathbf{y}^{\mathcal{C}_{\delta}}_{\mathcal{P}_1}=\mathbf{H}_I^{\dagger}\mathbf{W}_2\mathbf{x}_2+\mathcal{H}_1\mathbf{W}_1\mathbf{x}_1+\mathbf{n}_{\mathcal{P}_1},
\end{equation}
\begin{equation}
    \mathbf{y}^{\mathcal{C}_{\delta}}_{\mathcal{P}_2}=\mathbf{H}_h^{\dagger}\mathbf{W}_1\mathbf{x}_1+\mathcal{H}_2\mathbf{W}_2\mathbf{x}_2+\mathbf{n}_{\mathcal{P}_2},
\end{equation}
where $\mathbf{x}_1\in \mathbb{C}^{M_h\times 1}$ and $\mathbf{x}_2\in \mathbb{C}^{N_I \times 1}$ is the message signal vector of $\mathcal{P}_1$ and $\mathcal{P}_2$, respectively, and 
$\mathbf{n}_{\mathcal{P}_1}\in  \mathbb{C}^{M_I \times 1}$ and $\mathbf{n}_{\mathcal{P}_2}\in  \mathbb{C}^{N_h \times 1}$ are the additive white Gaussian noise (AWGN) vectors at the inputs of antennas at $\mathcal{P}_1$ and $\mathcal{P}_2$, respectively.
For the proposed communication setup, there are two performance metrics, namely, {\em instantaneous IT rate} at the $\mathcal{P}_2\xrightarrow{ }\mathcal{P}_1$ link and {\em energy transmission rate} at the $\mathcal{P}_1\xrightarrow{ }\mathcal{P}_2$ link. These two metrics can be expressed as 
\begin{dmath}
    R_{\mathcal{P}_1}^{\mathcal{C}_{\delta}}=\log\left |\mathbf{I}_{M_I} + \left(
 \mathbf{\dot{\Sigma}}_1+ \mathbf{\dot{\mathcal{H}}}_1\mathbf{W}_1\mathbf{\mathcal{R}}_{1}\mathbf{W}_1^{\dagger}\mathbf{\dot{\mathcal{H}}}_1^{\dagger} 
 \right)^{-1} \mathbf{H}_I \mathbf{W}_2\mathcal{R}_{2}\mathbf{W}_2^{\dagger} \mathbf{H}_I^{\dagger}\right|,\label{Rate_1}
\end{dmath}
\begin{dmath}
  \mathbb{E}\left[{\mathbf{y}^{\mathcal{C}_{\delta}}_{\mathcal{P}_2}}^{\dagger}\mathbf{y}^{\mathcal{C}_{\delta}}_{\mathcal{P}_2}\right]=\text{Tr}\left(
  \mathbf{H}_h\mathbf{W}_1\mathcal{R}_1\mathbf{W}_1^{\dagger}\mathbf{H}_h^{\dagger}+\ddot{\mathcal{H}}_2\mathbf{W}_2\mathcal{R}_2\mathbf{W}_2^{\dagger}\ddot{\mathcal{H}}_2^{\dagger}+\mathbf{\Sigma}_2
  \right),\label{Energy_1}
\end{dmath}
where $\mathbf{\mathcal{R}}_1\in \mathbb{C}^{M_h\times M_h}$ and $\mathbf{\mathcal{R}}_2\in \mathbb{C}^{N_I\times N_I}$ are the auto-correlation matrices of the transmitted signal defined as    $\mathbf{\mathcal{R}}_1=\mathbb{E}\left[\mathbf{x}_1\mathbf{x}_1^{\dagger} \right]$ and $\mathbf{\mathcal{R}}_2=\mathbb{E}\left[\mathbf{x}_2\mathbf{x}_2^{\dagger} \right]$, respectively, such that $\text{Tr}\left(\mathcal{R}_1\right)=P_{\text{S}}$, and $\text{Tr}\left(\mathcal{R}_2\right)\leq P_{\text{h}}$ where $P_{\text{S}}$ is the maximum power budget at $\mathcal{P}_1$,
$P_{\mathcal{\text{h}}}$ is the available power at $\mathcal{P}_2$ harvested from $\mathcal{P}_1$ during the previous time slot such that $0\leq P_{\text{h}}\leq P_{\text{S}}$, i.e. $P_h= \mathbb{E}\left[{\mathbf{y}^{\mathcal{C}_{\delta}}_{\mathcal{P}_2}}(t-1)^{\dagger}\mathbf{y}^{\mathcal{C}_{\delta}}_{\mathcal{P}_2}(t-1)\right]$.  $\mathbf{\Sigma}_1 \in \mathbb{C}^{M_I \times M_I}$ and $\mathbf{\Sigma}_2\in \mathbb{C}^{N_h \times N_h}$ are the auto-correlation matrices of the AWGN components at $\mathcal{P}_1$ and $\mathcal{P}_2$, respectively.

Accordingly, for the proposed system, the general antenna antenna allocation and precoding (or power allocation) problem  can be formulated as
\begin{equation}
\begin{aligned}
&  \textbf{P}_1~\text{:}\underset{{\delta}, \bm{W}_1, \bm{W}_2 }{\text{max}}  \text{\hspace{-0mm}} \alpha R_{\mathcal{P}_1}^{\mathcal{C}_{\delta}}+(1-\alpha)\mathbb{E}\left[{\mathbf{y}_2^{\mathcal{C}_{\delta}}}^{\dagger}\mathbf{y}_2^{\mathcal{C}_{\delta}}\right]
\\
& \text{\hspace{0mm} Subject to:} 
\\
&\text{\hspace{0mm}}\textbf{C}_1:\text{\hspace{-0mm}}\text{Tr} \left(\mathbf{W}^{ }_1\mathbf{\mathcal{R}}_1\mathbf{W}_1^{\dagger}\right)\leq P_{\mathcal{\text{S}}}\\
&\text{\hspace{0mm}}\textbf{C}_2:\text{\hspace{-0mm}}\text{Tr} \left(\mathbf{W}^{ }_2\mathbf{\mathcal{R}}_2\mathbf{W}_3^{\dagger}\right)\leq P_{\mathcal{\text{h}}}\\
&\text{\hspace{0mm}} \textbf{C}_3:\text{\hspace{-0mm}}   0\preceq\mathbf{W}_i \preceq 1, \forall i=1, 2~~\text{and}~~\delta=\{1, \dots, \Delta \} \label{GeneralProb}
\end{aligned}
\end{equation}
where $0<\alpha<1$ is a trade-off factor used to bias one of the two operations (IT or EH) over the other. Note that $\alpha$ cannot be 0 or 1 since both operations depend on the occurrence of the other one. The constraints $\textbf{C}_1$ and $\textbf{C}_2$ are related to maximum power budgets at $\mathcal{P}_1$ and $\mathcal{P}_2$, respectively.

Problem $\textbf{P}_1$ in (\ref{GeneralProb}) a combinatorial optimization problem. 
The complexity of this problem will increase exponentially with the number of antennas (and hence the possible MIMO subsystem configurations). This is due to the fact that the number of possible antenna allocation configurations can be considered as a modified version of the $n$-th order Bell number that is represented by sum of exponential functions (Dobinski's formula).  

For every possible antenna allocation configuration, there is  an optimal MIMO-FD precoding scheme. The globally optimal solution is the MIMO subsystem-precoding scheme pair that gives the best performance (i.e. the highest IT rate). This can only be achieved through optimizing the FD precoding scheme for every possible MIMO subsystem configuration, and therefore, an exhaustive search will be required. This is infeasible  for real-time implementation at the  devices with limited battery and processing capabilities.  

In the following section, we will propose a sub-optimal solution that first splits the general optimization problem into two sub-problems: one for antenna allocation and the other for precoding matrix design.  Then, we  solve the sub-problems separately.


\section{Sub-Optimal Antenna Allocation and Relaxation-Based Precoding}
\subsection{Antenna Allocation Method}

We present a sub-optimal scheme that allocates antennas between EH and IT with reduced computational complexity.  
When designing an antenna allocation scheme for the proposed system, the following parameters need to be taken into consideration:
\begin{itemize}  
    \item The rank of both IT and EH channel matrices. This is due to the fact that any MIMO system can achieve up to $K$-fold increase in transmission capacity (or harvested energy), where $K$ is the minimum number of antennas between MIMO transmitter and receiver. 
    Accordingly, any antenna allocation method should split antennas such that the two MIMO subsystems have the maximum possible rank.
    \item One way to maximize the rank of both IT and EH matrices is by selecting by splitting the columns (or rows) pairs of $\bm{H}$ with the highest correlation (highest dot product) between IT and EH subsystems.
    \item Note that the maximum energy that can be harvested by $\mathcal{P}_2$ is assumed to be equal to the maximum power budget of $\mathcal{P}_1$. Accordingly, the number of antennas allocated for EH should be monitored such that no extra harvested power is discarded in a certain time slot.  
\end{itemize}
In \textbf{Algorithm \ref{Algorithm1}}, we propose a  procedure that achieves a balance between increasing the amount of harvested energy at $\mathcal{P}_2$ and the amount of information received at $\mathcal{P}_1$.
\begin{algorithm}
  \caption{\textbf{\hspace{-1.5mm}}: Antenna allocation at
$\mathcal{P}_1$ and $\mathcal{P}_2$
}\label{Algorithm1}
  \begin{algorithmic}[1]
        \State \textbf{\textit{Input:}} $\mathbf{H}$, $P_S$ and $P_Q$  \State \textbf{\textit{Initialize:}} $\Psi_{1,\text{I}}=\{1, \dots, M\}$, $\Psi_{2,\text{I}}=\{1, \dots, N\}$,  $\Psi_{1,\text{h}}=\Psi_{2,\text{h}}=\Phi, M_I=M$ and $N_I=N$.
        \State $(m^*,n^*)=\min_{m,n}||h_{n,m}||^2$
        \State $\Psi_{1,h}=\Psi_{1,h}\cup m^*~\&~ \Psi_{1,I}=\Psi_{1,I}\backslash m^*$
        \State $\Psi_{2,h}=\Psi_{2,h}\cup n^*~\& \Psi_{2,I}=\Psi_{2,I}\backslash n^*$
      \While{ $ \sum_{n \in \Psi_{2,h}}^{ } \sum_{m \in \Psi_{1,h}}^{ } \frac{P_S |h_{n,m}|^2}{M_h} < P_Q~\&~M_I>1~\&~N_I>1$ }
        \State \textbf{Find:} $\textbf{z}_1=||\mathbf{H(:,\Psi_{1,I})}||_2$ and  $\textbf{z}_2=||\mathbf{H(\Psi_{2,I},:)}||_2$
        \State \textbf{Find:} $m^*=\min_{m\in \Psi_{1,I}}\textbf{z}_1$ and $n^*=\min_{n\in \Psi_{2,I}}\textbf{z}_2$ 
        \If{$\mathbf{Z}_1(m^*)\leq \mathbf{Z}_2(n^*) $}
$\Psi_{1,h}=\Psi_{1,h}\cup m^*~\& \Psi_{1,I}=\Psi_{1,I}\backslash m^*$
    \EndIf
    \If{$ \mathbf{Z}_2(n^*)<\mathbf{Z}_1(m^*) $}
$\Psi_{2,h}=\Psi_{2,h}\cup n^*~\& \Psi_{2,I}=\Psi_{2,I}\backslash n^*$
    \EndIf
      \EndWhile\label{euclidendwhile}
      \State \textbf{Return} $\Psi_{1,I}$, $\Psi_{2,I}$, $\Psi_{1,h}$ and $\Psi_{2,h}$
  \end{algorithmic}
\end{algorithm}
In this algorithm, $\Psi_{i,I}$ and $\Psi_{i,h}, i=1,2$ are sets containing the indices of antennas used for IT and EH, respectively, at $\mathcal{D}_i$. $P_Q$ is a design parameter that specifies the approximate amount of energy to be harvested for a given antenna allocation.    
After antenna allocation, the P2P system will split up into two virtual parallel MIMO subsystems: one for EH and the other for IT with channel matrices denoted by $\mathbf{H}_I\in~~\mathbb{C}^{M_I \times N_I}$ and $\mathbf{H}_h\in~~\mathbb{C}^{N_h \times M_h}$, respectively. 

\subsection{Relaxation-Based MIMO Precoding Design}
After finding the MIMO subsystems based on the proposed antenna allocation algorithm (\textbf{Algorithm 1}), 
let $\mathbf{Q}_1=\mathbf{W}_1\mathcal{R}_1\mathbf{W}_1^{\dagger}$ and $\mathbf{Q}_2=\mathbf{W}_2\mathcal{R}_2\mathbf{W}_2^{\dagger}$.
Note that, since $\mathcal{R}_1$ and $\mathcal{R}_2$ are Hermitian (auto-correlation) matrices and $\mathbf{W}_i\in \mathcal{R}^+, i=1, 2$,  $\mathbf{Q}_i$s are Hermitian positive-definite matrices whose Cholesky decomposition can be written as $\mathbf{Q_i}=\mathbf{B}_i^{ }\mathbf{B}_i^{\dagger}, i=1, 2$, where $\mathbf{B}_i$ is a lower triangular matrix. Accordingly, the performance metrics in (\ref{Rate_1}) and (\ref{Energy_1}) reduce to
\begin{equation}
    R_{\mathcal{P}_1}^{\mathcal{C}_{\delta^*}}=\log\left |\mathbf{I}_{M_I} + \left(
 \mathbf{\dot{\Sigma}}_1+ \mathbf{\dot{\mathcal{H}}}_1\mathbf{Q}_1\mathbf{\dot{\mathcal{H}}}_1^{\dagger} 
 \right)^{-1} \mathbf{H}_I\mathbf{Q}_2 \mathbf{H}_I^{\dagger}\right|,\label{Rate_2}
\end{equation}
\begin{equation}
  \mathbb{E}\left[{{\mathbf{y}_{2}^{\mathcal{C}_{\delta^*}}}^{\dagger}\mathbf{y}_{2}^{\mathcal{C}_{\delta^*}}}\right]=\text{Tr}\left(
  \mathbf{H}_h\mathbf{Q}_1\mathbf{H}_h^{\dagger}+\dot{\mathcal{H}}_2\mathbf{Q}_2\dot{\mathcal{H}}_2^{\dagger}+\mathbf{\dot{\Sigma}}_2
  \right).\label{Energy_2}
\end{equation}
where ${\mathcal{C}_{\delta^*}}$ is the $\delta^*$-th MIMO subsystem configuration selected by using \textbf{Algorithm 1}. 


When the problem in (\ref{GeneralProb}) is divided into two separate sub-problems (i.e. MIMO antenna allocation problem and precoding design problem),  the non-linearity of the antenna allocation constraints is eliminated. However, even for a given antenna allocation, the precoding problem is non-convex due to the fact that $R_{\mathcal{P}_1}^{\mathcal{C}_{\delta}}$ represents a non-convex function of $\mathbf{W}_1$ and $\mathbf{W}_2$. Therefore, a globally optimal solution for this sub-problem can be obtained only by an exhaustive search.

To further proceed to obtain a practical solution to this problem, let us first rewrite (\ref{Rate_2}) as follows:

\begin{dmath}
R_{\mathcal{P}_1}^{\mathcal{C}_{\delta^*}}= \log \left|
\mathbf{\dot{\Sigma}}_1+ \dot{\mathcal{H}}_1\mathbf{Q}_1\dot{\mathcal{H}}_1^{\dagger}
+
\mathbf{{H}}_I\mathbf{Q}_2\mathbf{{H}}_I^{\dagger}
\right|
-
\log\left|
\mathbf{\dot{\Sigma}}_1+ \dot{\mathcal{H}}_1\mathbf{Q}_1\dot{\mathcal{H}}_1^{\dagger}
\right|.\label{Rate3}
\end{dmath}
Note that the right hand side of (\ref{Rate3}) can be approximated by a concave function if we are able to approximate the second term (let us denote it by $f\left(\mathbf{Q}_1 \right)=\log\left|
\mathbf{\dot{\Sigma}}_1+ \dot{\mathcal{H}}_1\mathbf{Q}_1\dot{\mathcal{H}}_1^{\dagger}
\right|$) by an affine function. 
We utilize the fact that the Taylor polynomial with the first two terms in the Taylor series expansion of $R_{\mathcal{P}_1}^{\mathcal{C}_{\delta}}\left(\mathbf{Q}_1\right)$ is an affine function that represents a global under-estimator of $R_{\mathcal{P}_1}^{\mathcal{C}_{\delta}}\left(\mathbf{Q}_1\right)$ \cite[Eq. 3.2]{Boyd2004}. 
In other words,\footnote{Note that (\ref{Taylor}) is the first order Taylor approximation for a scalar function of a vector. In our case, where we generally deal with scalar function of a matrix ($\mathbf{Q}_1$), we need to apply the definition directly and rewrite the results in a canonical matrix form.}
  
\begin{equation}
    f\left(\mathbf{Q}_1\right)\approx f\left(\mathbf{Q}_1^{o}\right)+
    \nabla f\left(\mathbf{Q}_1^{o}\right)^{\dagger}\text{Vec}\left(\mathbf{Q}_1-\mathbf{Q}_1^{o} \right)\label{Taylor},
\end{equation}
where $\mathbf{Q}_1^{o}\geq \mathbf{0}$ is the matrix at which $R_{\mathcal{P}_1}^{\mathcal{C}_{\delta}}$ is expanded around an arbitrary operating point.
Applying ($\ref{Taylor}$) to the second element of (\ref{Rate3}) and utilizing the fact that $d \log |f(\mathbf{X})|=\text{Tr}\left(f(\mathbf{X})^{-1}df(\mathbf{X})\right)$ and $\text{Tr}\left(\mathbf{I}_{M_I}^{\dagger}\left(\mathbf{Q}_1-\mathbf{Q}_1^o\right)\right)=\text{Vec}\left( \mathbf{I}_{M_I}\right)^{\dagger}\text{Vec}\left( \mathbf{Q}_1-\mathbf{Q}_1^o\right)$, $R_{\mathcal{P}_1}^{\mathcal{C}_{\delta}}$ can be written as (\ref{Linear}). 

Note that the effect of removing the higher order terms in (\ref{Taylor}) will be significant when the operating points (transmission powers) of the system lie in the curvature vicinity of the transmission rate function. This is due to the fact that (\ref{Taylor}) is a linear approximation that represents a tangent line with the exact $R_{\mathcal{P}_1}^{\mathcal{C}_{\delta}}$ at $\mathbf{Q}_1^o$. 
\begin{figure*}[!t]
\normalsize
 \begin{equation}
    \begin{aligned}
&\text{\hspace{-46mm}}R_{\mathcal{P}_1}^{\mathcal{C}_{\delta^*}}=
\log\left |\mathbf{I}_{M_I} + \left(
 \mathbf{\dot{\Sigma}}_1+ \mathbf{\dot{\mathcal{H}}}_1\mathbf{Q}_1\mathbf{\dot{\mathcal{H}}}_1^{\dagger} 
 \right)^{-1} \mathbf{H}_I\mathbf{Q}_2 \mathbf{H}_I^{\dagger}\right|\\
&\text{\hspace{-38mm}} = \log \left|
\mathbf{\dot{\Sigma}}_1+ \dot{\mathcal{H}}_1\mathbf{Q}_1\dot{\mathcal{H}}_1^{\dagger}
+
\mathbf{{H}}_I\mathbf{Q}_2\mathbf{{H}}_I^{\dagger}
\right|
-
\log\left|
\mathbf{\dot{\Sigma}}_1+ \dot{\mathcal{H}}_1\mathbf{Q}_1\dot{\mathcal{H}}_1^{\dagger}
\right|\\
&\text{\hspace{-38mm}} \approx \log \left|
\mathbf{\dot{\Sigma}}_1+ \dot{\mathcal{H}}_1\mathbf{Q}_1\dot{\mathcal{H}}_1^{\dagger}
+
\mathbf{{H}}_I\mathbf{Q}_2\mathbf{{H}}_I^{\dagger}
\right|
+\frac{1}{\ln{2}}\text{Tr}
\left[
\left(\dot{
\mathbf{\Sigma}}_1+ \dot{\mathcal{H}}_1\mathbf{Q}_1^{o}\dot{\mathcal{H}}_1^{\dagger}
\right)^{-1}
\dot{\mathcal{H}}_1\left(\mathbf{Q}_1^{o}-\mathbf{Q}_1\right)\dot{\mathcal{H}}_1^{\dagger}
\right]
-
\\
\text{\hspace{10mm}}
\log\left|
\mathbf{\dot{\Sigma}}_1+ \dot{\mathcal{H}}_1\mathbf{Q}_1^{o}\dot{\mathcal{H}}_1^{\dagger}
\right|.
    \end{aligned}\label{Linear}
\end{equation}
\hrulefill
\vspace*{6pt}
\end{figure*}

Now that we have changed the objective function to an affine function, the general optimization problem can be expressed as
\begin{equation}
\begin{aligned}
& \text{\hspace{4mm}} \underset{\mathbf{Q}_1, \mathbf{Q}_2 }{\text{max}}
&& \alpha R_{\mathcal{P}_1}^{\mathcal{C}_{\delta^*}} 
+
(1-\alpha)\mathbb{E}\left[{\mathbf{y}^{\mathcal{C}_{\delta^*}}_2}^{\dagger}\mathbf{y}^{\mathcal{C}_{\delta^*}}_2\right]
\\
& \text{\hspace{4mm} Subject to:} 
&&\text{C}_1: \text{\hspace{-0mm}}\text{Tr} \left(\textbf{Q}_1\right)\leq P_{\mathcal{\text{S}}}\\
&&&\text{C}_2:\text{\hspace{-0mm}}\text{Tr} \left(\textbf{Q}_2\right)\leq P_{\mathcal{\text{Q}}}\\
&&& \text{C}_3:\text{\hspace{-0mm}}   \mathbf{Q}_i \succeq 0, \forall i=1, 2. \label{GeneralProb2}
\end{aligned}
\end{equation}

After solving (\ref{GeneralProb2}) for $\mathbf{Q}_1$ and $\mathbf{Q}_2$, the precoding matrices can be easily found by solving $\mathbf{w}_i=\mathbf{B}_i\mathcal{R}_i^{-\frac{1}{2}}, i=1, 2$, where $\mathbf{B}_i$ is the Cholesky decomposition as of $\mathbf{Q}_i$ as shown before.
This problem can be easily solved by using convex optimization algorithms available in optimization tool boxes such as CVX in Matlab. 

A wise selection of $\mathbf{Q}_1^{o}$ plays a major role in determining the accuracy of the proposed solution and the corresponding approximation.
This is due to the fact that the approximation in (\ref{Taylor}) is accurate only for a certain region that is centred around $\mathbf{Q}_1^{o}$ and the accuracy decreases as we go fare from $\mathbf{Q}_1^o$.
This may lead to a selection of $\mathbf{Q}_1^*$ and $\mathbf{Q}_2^*$ that optimize another objective function rather than $\alpha R_{\mathcal{P}_1}^{\mathcal{C}_{\delta^*}}+(1-\alpha)\mathbb{E}\left[{\mathbf{y}_1^{\mathcal{C}_{\delta^*}}}^{\dagger}\mathbf{y}^{\mathcal{C}_{\delta^*}}_1\right]$. 
In \cite{6990627}, the authors proposed that $\mathbf{Q}_1^{o}$ should be first selected randomly and then updated based on the solution of the optimization problem until the algorithm converges (i.e. there is no more enhancement on the original objective function).
Here, we propose a modified algorithm (\textbf{Algorithm \ref{Algorithm2}}) to solve the problem in (\ref{GeneralProb2}). 

\begin{algorithm}
  \caption{Precoding algorithm}\label{Algorithm2}
  \begin{algorithmic}[1]
        \State \textbf{\textit{Input:}} $\mathbf{H}_h$, $\mathbf{H}_I$, $\dot{\mathcal{H}}_1$, $\dot{\mathcal{H}}_2$,
        $\mathcal{R}_1$,
        $\mathcal{R}_2$
        $\mathbf{\dot{\Sigma}}_1$, $\mathbf{\dot{\Sigma}}_2$, $P_S$ and $\alpha$. 
      \State \textbf{\textit{Initialize:}} $\mathbf{Q}_1^{o}=\frac{P_S}{M_h}\mathbf{I}_{M_h}$\Comment{Equal Power Allocation}
      \While{$\alpha \,R_{\mathcal{P}_1}^{\mathcal{C}_{\delta}} 
+
(1-\alpha)\,\mathbb{E}\left[{\mathbf{y}_1^{\mathcal{C}_{\delta}}}^{\dagger}\mathbf{y}^{\mathcal{C}_{\delta}}_1\right]$ "not converged"}
        \State Solve (\ref{GeneralProb2}) to obtain $\mathbf{Q}_1^*$ and $\mathbf{Q}_1^*$.
        \State Set: $\mathbf{Q}_1^o=\mathbf{Q}_1^*$
        \State Update $R_{\mathcal{P}_1}$ in (\ref{Rate3}) with new $\mathbf{Q}_1^o$
      \EndWhile\label{euclidendwhile}
      \State Find $\mathbf{B}_1$ and $\mathbf{B}_2$ s.t. $\mathbf{Q}_i^*=\mathbf{B}_i\mathbf{B}_i^{\dagger}$
      \State $\mathbf{W}_i=\mathbf{B}_i \mathcal{R}^{-\frac{1}{2}}_i, i=1, 2.$
  \end{algorithmic}
\end{algorithm}

As can be seen in \textbf{Algorithm \ref{Algorithm2}}, we first approximate the non-linear part of $R_{\mathcal{P}_1}$ around $\mathbf{Q}_1^o$ that is corresponding to equal power allocation constraints. This is reasonable to assume that under i.i.d. channel conditions, the optimal solution varies slightly  from that due to equal power allocation. 

\section{Deep Reinforcement Learning-Based Solution}
In the previous section, we proposed a  MIMO antenna allocation algorithm that finds the MIMO subsystems for EH and IT followed by an approximate MIMO precoding scheme.
However, the optimal solution for the joint antenna allocation and MIMO precoding for the proposed system can only be obtained by testing every possible antenna allocation configuration, finding the corresponding optimal precoding matrices ($\bm{W}_1$ and $\bm{W}_2$), and then choosing the solution that results in the best performance. Such an exhaustive search-based solution will incur a significant amount of processing complexity and delay which will make the proposed system impractical. In this section, we propose a novel DRL model that jointly performs antenna allocation scheme and MIMO precoding at the two devices $\mathcal{P}_1$ and $\mathcal{P}_2$.

Fig. \ref{AI_Model} describes the high-level architecture of the DRL framework proposed in this paper. The framework consists of a  simulated MIMO-P2P-FD environment as shown in Fig. \ref{SystemModel} along with a learning agent. Before delving into the specifics of the DRL model, we first introduce the DRL preliminaries and the associated terminologies.

\subsection{Deep Reinforcement Learning}
Reinforcement learning (RL) is a learning approach that falls between supervised and unsupervised learning. It is neither strictly supervised as it does not rely on a labeled data set, nor fully unsupervised  since the learning agent uses a reward signal provided by an environment. In the RL setting, the agent aims to select the right action for the next interaction in order to maximize the discounted reward over a finite time horizon. This problem is commonly formulated as a \textit{Markov Decision
Process} (MDP) problem. An MDP is a tuple ($\mathcal{\bm S}$, $\mathcal{A}$, $\mathcal{P}$, $\mathcal{R}$, $\zeta$), where $\mathcal{S}$ is the state space that consists of the set of all possible $K$-dimensional states, $\mathcal{A}$ is a finite set of actions from which the agent can choose, $\mathcal{P}$ : $\mathcal{S}$ $\times$ $\mathcal{A}$ $\times$ $\mathcal{S}$ $\rightarrow$ [0, 1] is a transition probability in which $\mathcal{P}(\bm{s}, a, \bm{s}')$ defines the probability of observing state $\bm{s}'$ after executing action $a$ in the state $\bm{s}$, $\mathcal{R}: \mathcal{S} \times \mathcal{A} \rightarrow \mathbb{R}$ is the expected reward after being in state $\bm{s}$ and taking action $a$, and $\zeta$ $\in [0, 1)$ is the discount factor.

Developing RL algorithms to solve an MDP problem requires finding a discrete value function or a ``policy" which maps the  observations to the next action to be taken by the agent. Opting for the discretization of the policy's action space can lead to a lack of generalization and significantly increases the problem's dimensionality.
Therefore, deep RL (DRL) algorithms based on function approximation by deep neural networks (DNNs) have been proposed.

DRL algorithms can be categorized into three types: (i) \textit{value-based} methods such as Q-learning and SARSA that learn the so-called  value function to find a policy, (ii) \textit{policy-based} methods which learn the policy directly by following its gradient, and (iii) \textit{actor-critic} methods that combine the value-based and the policy-based methods where the policy is known as the actor, because it is used to select actions, and the estimated value function is known as the critic, because it criticizes the actions made by the actor. 

The standard Q-Learning (QL) method represents the most popular update algorithm in the RL literature. The QL update equation at time $t$ for a network agent with parameters $\theta^Q$ after taking action $a_t$ in state $s_t$ and observing the
immediate reward $r_{t+1}$ and resulting state $s_{t+1}$ is:
\begin{equation}
\label{eq:dqn-update equation}
\begin{aligned}
Q(s, a \,|\, \theta^Q_{t+1}) &= Q(\bm{s}, a\,|\, \theta^Q_t) + \nu \bigg[r_{t+1} + \zeta  \max_{a'} Q(\bm{s}_{t+1}, a'\,|\, \theta^Q_t) - Q(\bm{s}_t, a_t\,|\, \theta^Q_t)\bigg]\\
&\text{\hspace{-30mm}}= Q(\bm{s}, a\,|\, \theta^Q_t) + \nu \bigg[r_{t+1} + \zeta \max_{a'} Q(\bm{s}_{t+1}, \argmax_{a'} Q(s_{t+1},a'\,|\, \theta^Q_t)\,|\, \theta^Q_t) - Q(\bm{s}_t, a_t\,|\, \theta^Q_t)\bigg],
\end{aligned}
\end{equation}
where $\nu$ is the learning rate.
 The Q-learning update in (\ref{eq:dqn-update equation}) overestimates the Q-values because it uses the same Q-network $Q(\bm{s}, a\,|\, {\theta}^Q_t)$ both to select and to evaluate an action. Decoupling the action selection and evaluation steps avoids the maximization bias. Hasselt et al. \cite{van2016deep} introduced the \textit{Deep Double Q-Learning} algorithm (DDQL) that uses two deep Q-networks: a  $Q$ network and a target network $Q'$ with different parameters $\theta^Q$ and ${\theta}^{Q'}$, respectively, to achieve such decoupling. The $Q$ network is used for action evaluation while the  $Q'$ network is used for action selection.
 
The DDQL update equation of the network can be expressed as:
\begin{dmath}
\text{\hspace{-2mm}}Q(\bm{s}, a \,|\, {\theta}^Q_{t+1}) = Q(\bm{s}, a\,|\, {\theta}^Q_t) 
\\
+ \nu \bigg[r_{t+1} + \zeta \max_{a'} Q(\bm{s}_{t+1}, \argmax_{a'} Q'(\bm{s}_{t+1},a'\,|\, {\theta}^{Q'}_t)\,|\, {\theta}^Q_t) - Q(\bm{s}_t, a_t\,|\, {\theta}^Q_t)\bigg].
\label{eq:ddqn-update-q1-equation}
\end{dmath}
The parameters ${\theta'}^Q$ of the  $Q'$ network periodically hard-copy the parameters $\theta^Q$ of $Q$ network after $t_0$ time steps using the Polyak averaging method with parameter $\tau \in [0,1]$:
\begin{equation}
\label{eq:ddqn-update-q2-equation}
{\theta}^{Q'}_{t+t_0} = (1-\tau) \,{\theta}^{Q'}_t + \tau \,\theta^Q_t.
\end{equation}

DDQL achieves a better performance than standard DQL \cite{van2016deep}; however, due to the discretization requirements of the DNN outputs (the action space $\mathcal{A}$), it results in a huge expansion of the action space dimensionality when used for optimization of an objective function of continuous dependent variables.
{\em This dimensionality issue makes it an unattractive solution for solving the beamforming problem under a large number of antennas at the two devices $\mathcal{P}_1$ and $\mathcal{P}_2$.}
However, it is a relevant candidate for the antenna allocation problem  since it avoids the need for an extremely inefficient exhaustive search method. 
This  motivates us to utilize the ``DDPG" policy for the precoding design problem.

The DDPG is an actor-critic algorithm. It concurrently learns a policy network approximation $\mu(\bm{s}|\theta^{\mu})$ called the actor, and a Q-function network approximation $Q(\bm{s}, a|\theta^{Q})$ called the critic. The Q-function network is trained using the Bellman equation, while the policy network is learnt using the Q-function. The output of the policy network of DDPG directly maps the states to actions, instead of computing the probability distribution $\pi(a|\bm{s})$ across a discrete action space $\mathcal{A}$ which is the case of the DQL policies. At every time step $t$, the policy maximizes its loss function defined as:
\begin{equation}
\label{eq:loss-function-actor}
J(\theta) = \mathbb{E}\bigg[Q(\bm{s}, a) \;|\; \mathcal{S}=\bm{s}_t, a=\pi(a|\bm{s}_t)\bigg]
\end{equation}
and updates its weights $\theta$ by following the gradient of (\ref{eq:loss-function-actor}):
\begin{equation}
\label{eq:gradient-loss-function-actor}
\nabla J_{\theta^{\mu}}(\theta) \approx \nabla_{a} Q(\bm{s},a) \,\nabla \mu(s|\theta^{\mu}).
\end{equation}
This update rule represents the deterministic policy gradient (DPG) theorem, rigorously proved by Silver et al. in the supplementary material of \cite{silver2014deterministic}.
The term $\nabla_{a} Q(\bm{s},a)$ is obtained from a Q-network  $Q(\bm{s}, a|\theta^{Q})$ called the critic by backpropagating its output w.r.t. the action input $\mu(\bm{s}|\theta^{\mu})$. When the number of actions is very large, this actor-critic training procedure solves the intractability problem of DQN \cite{mnih2015human}  by using the following approximation:
\begin{equation}
\label{eq:max-approx-ddpg}
\max_{a} Q(\bm{s}, a) \approx Q(\bm{s}, a |\theta^{Q})|_{a=\mu(\bm{s}|\theta^{\mu})}.
\end{equation}
Similar to DQN, two tricks are employed to stabilize the training of the DDPG actor-critic architecture: i) the experience replay buffer $R$ to train the critic, and ii) target networks for both the actor and the critic which are updated using the Polyak averaging in the same way it was done in (\ref{eq:ddqn-update-q2-equation}).
In the following subsection, we provide a detailed description of the proposed DRL-based antenna allocation and precoding design for the MIMO-P2P-FD system.


\subsection{DRL Agent Design for Antenna Allocation and Precoding Optimization}

We design a DRL system that jointly optimizes the antenna allocation for EH and IT  and the precoding matrices at $\mathcal{P}_1$ and $\mathcal{P}_2$ given a certain CSI matrix
$\bm{H}$ similar to the one proposed in \cite{Cell_Free_Yasse}.
In this context, we develop a hybrid DDPG-DDQL DRL scheme that concurrently learns the best MIMO subsystem-precoding matrix pairs given a certain CSI matrix $\bm{H}$.

\begin{figure}[htb]
		\centering
		\includegraphics[scale=0.35]{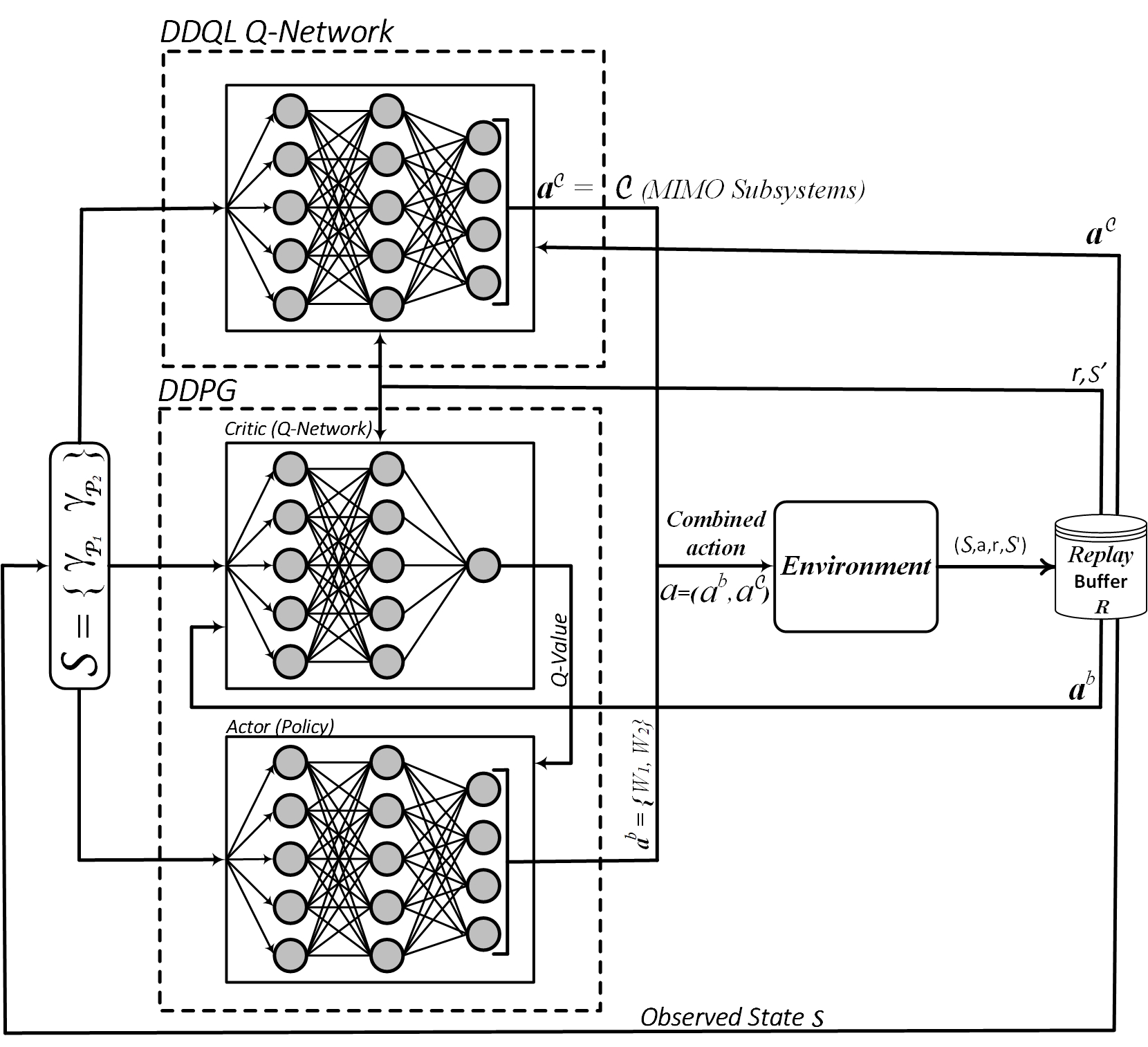}
		\caption{Hybrid DDPG-DDQL model for antennas' splitting and FD MIMO precoding. }\label{AI_Model}
	\end{figure}

Fig. \ref{AI_Model} shows a schematic block diagram of the developed hybrid DDPG-DDQL DRL model.
To find the MIMO precoding matrices $\bm{W}_1$ and $\bm{W}_2$, we first merge $\bm{W}_1$ and $\bm{W}_2$ into one vector, i.e. $\bm{w}=\left[\text{Span}\left(\bm{W}_1\right)~\text{Span}\left(\bm{W}_1\right)\right]^T$, where $\bm{w}\in \mathbb{R}^{M_h\times N_h + M_I\times N_I,1}$, and then use the actor-critic DDPG algorithm \cite{lillicrap2015continuous} since all the  elements of $\bm{w}$ are in the continuous range $[0,1]$.
For the antenna allocation problem, we use the well-known DDQL algorithm \cite{van2016deep} to find the best values of $M_I, M_h, N_I$, and $N_h$ and their corresponding MIMO subsystem configurations (the indices of antennas used for EH and IT). Antenna allocation based on the DDQL algorithm is justified by the fact that all possible MIMO subsystem configurations belong to a finite discrete set with integer indexing of every possible configuration.

Note that the two algorithms interact with a simulated MIMO-P2P-FD environment to solve the optimization problem $\textbf{P}_1$ in (\ref{GeneralProb}), i.e. simultaneous interactive learning is performed.
The design of the MIMO-P2P-FD environment involves the specification of the environment state $\bm{s}$ and the definition of the immediate reward function $r$ required by the DRL algorithms (DDPG and DDQN) to approximate the policies and the Q-values. 
The state of the environment is a vector
consisting of two elements, namely, the ${\rm SINR}$ value at $\mathcal{P}_2$ (denoted by $s_1=\gamma_{\mathcal{P}_1}^{\mathcal{C}_{\delta}}$) and the amount of energy harvested at $\mathcal{P}_2$ (denoted by $s_2=\mathbb{E}\left[{\bm{y}_{\mathcal{P}_2}^{\mathcal{C}_{\delta}} }^{\dagger}\bm{y}_{\mathcal{P}_2}^{\mathcal{C}_{\delta}}\right]$). These two states are a functions of current CSI ($\bm{H}$) and MIMO subsystem configuration $\mathcal{C}_{\delta}$.
Under this setup, the action space $\mathcal{A}$ is the pair of variable actions $a=(a^{p}, a^{\mathcal{C}}) = (\bm{w},\mathcal{C}_{\delta})$ that the DDPG and the DDQN output separately. 
The superscripts $p$ and $\mathcal{C}$ refer to  ``\textit{precoding}" and ``\textit{antenna allocation}", respectively. 
After receiving the environment's state $\bm{s}$, the DDPG algorithm outputs the action $a^{p} = \bm{w}$ and the DDQL outputs  the action $a^{\mathcal{C}}$ representing the $\delta$-th MIMO subsystem configuration $\mathcal{C}_{\delta}$. 


Table \ref{Table3} summarizes the environment design by specifying the additional problem parameters.
\begin{table}[h!]
    \centering
      \caption{DRL agent design}
\begin{tabular}{|c|c|}
\hline
Environment Variables & System Equivalence  \\
\hline
\hline
State $\mathcal{\bm{s}}=\{s_1, s_2 \}$ & \{$\mathbb{E}\left[{\mathbf{y}_{\mathcal{P}_2}^{\mathcal{C}_{\delta}}}^{\dagger}{\mathbf{y}_{\mathcal{P}_1}^{\mathcal{C}_{\delta}}} \right], \gamma_{\mathcal{P}_2}^{\mathcal{C}_{\delta}}\}$\\
\hline
Reward $r$ & $R_{\mathcal{P}_2}^{\mathcal{C}_{\delta}}= \log\left(1+\gamma_{\mathcal{P}_2}^{\mathcal{C}_{\delta}}\right)$\\
\hline
Action $\mathcal{A}$ & $(a^{p}, a^{\mathcal{C}}) = (\bm{w}, \mathcal{C}_{\delta}) = (\text{Span}([\bm{W}_1, \bm{W}_2 ]), \mathcal{C}_{\delta})$ \\
\hline
$\delta=1, \dots, \Delta$ & Index of possible MIMO subsystems configuration \\
\hline
$\mathcal{C}_{\delta}$ & $\delta$-th antennas' allocation configuration \\
\hline
\end{tabular}
    \label{Table3}
\end{table}
Our DRL implementation uses TensorFlow and Keras to train all the networks. We train DDQN and DDQL networks over 2500 episodes, with 500 time step each. The actor and critic networks and their corresponding targets, have two hidden layers of 256 and 128 neurons, respectively. The DDQN networks have two fully-connected layers of 64 neurons followed with the nonlinear activation function \textit{relu} each, and a final linear fully-connected layer. Our code uses a discount factor $\zeta = 0.99$, a learning rate $\nu = 5 \,\cdot\,10^{-5}$, a Polyak averaging parameter $\tau=10^{-3}$, and an experience replay buffer of size $R=20000$. The optimizer of the critic network is Adam with its default hyperparameters $\beta_1=0.9$ and $\beta_2=0.999$.

\subsection{Description of the Hybrid DDPG-DDQL Algoritm}
The neural networks used in \textbf{Algorithm \ref{algo:DRL-for-free-cell-network}} are concurrently trained by interacting with the MIMO-P2P-FD environment. In this section, we describe the role of every network and detail all the steps of the training process.
\begin{itemize}
    \item \underline{DDPG network}:
    \begin{itemize}
        \item \textit{The actor network $\mu ({\bm{s}} | \theta^{\mu})$} maps $s_1$ and $s_2$ values of devices $\mathcal{P}_1$ and $\mathcal{P}_2$ to the precoding vector $\bm{w}$. The output of the network is $a^{p}$, i.e. a flatten list of all the combined elements of $\bm{W}_1$ and $\bm{W}_2$.
        \item \textit{The target actor network $\mu' ({\bm{s}} | \theta^{\mu'})$}: time-delayed copy of the actor network $\mu ({\bm{s}} | \theta^{\mu})$.
        \item \textit{The critic network $Q({\bm{s}}, a^p | \theta^{Q})$}: maps $s_1$ and $s_2$ values and the output action of $\mu ({\bm{p}} | \theta^{\mu})$ to their corresponding Q-value.
        \item \textit{The target critic network $Q'({\bm{s}}, a^p | \theta^{Q'})$}: time-delayed copy of the critic network $Q({\bm{s}}, a^p | \theta^{Q})$.
    \end{itemize}
    \item \underline{DDQL network}:
    \begin{itemize}
        \item \textit{The $Q_c$-network $Q_c({\bm{s}}, a^{\mathcal{C}} | \theta^{Q_c})$} maps $s_1$ and $s_2$ values of devices $\mathcal{P}_1$ and $\mathcal{P}_2$ to the Q-values of the state and all the possible antenna allocation partitions.
        \item \textit{The target $Q_c$-network $Q_c'({\bm{s}}, a^{\mathcal{C}} | \theta^{Q_c'})$}: time-delayed copy of the $Q_c$-network $Q({\bm{s}}, a^{\mathcal{C}} | \theta^{Q_c})$.
    \end{itemize}
\end{itemize}

We describe all of the training steps in \textbf{Algorithm \ref{algo:DRL-for-free-cell-network}}. In lines \ref{algo:init1}--\ref{algo:init5}, we begin by initializing all neural networks and their corresponding targets for the antenna allocation and MIMO precoding as well as a replay buffer $R$.
For every episode, we initialize the $M\times N$ MIMO system environment by first assuming an initial MIMO subsystem configuration $\mathcal{C}_o$ (chosen randomly) and equal power allocation precoding matrices $\bm{W}_1$ and $\bm{W}_2$ which gives an initial state $\bm{s}_0=\{s_1^o,s_2^o \}$ (line \ref{algo:init-env}). 

At every time step $t$ of the episode, the DDQL and DDPG agents pick an action $a^{\mathcal{C}}_t$ and $a^p_t$, respectively (lines \ref{algo:pick-ab}--\ref{algo:pick-ac}). The combined action $a_t = (a^p_t, a^{\mathcal{C}}_t)$ is sent to the MIMO-P2P-FD environment which will transit to a new state $\bm{s}_{t+1}$. This new state will be returned together with the immediate reward $r_t$ (lines \ref{algo:combine-ab-ac}--\ref{algo:get-state-reward}). After storing the transition tuple $(\bm{s}_{t}, a_{t}, r_{t}, \bm{s}_{t+1})$ in the replay buffer $R$ (line \ref{algo:store-transition}), we randomly sample from the experience replay buffer $N$ transitions to train the DDPG and DDQL networks (line \ref{algo:sample-replay-buffer}). 

We start the DDPG training in line (\ref{algo:ddpg-td-target}) by computing the target for the Q-network $Q(\bm{s}, a | \theta^{Q})$ using the target Q-network $Q'(\bm{s}, a | \theta^{Q'})$. We update the critic $Q(\bm{s}, a | \theta^{Q})$'s parameters $\theta^{Q}$ in line \ref{algo:update-critic} using the gradient of the mean squared error of the loss function of the target and the output of the critic. The update of the actor's parameters $\theta^{\mu}$ uses the Monte Carlo approximation of gradient in line \ref{eq:gradient-loss-function-actor}. 
The target critic and target policy networks are updated slowly every $P$ iterations (lines \ref{algo:update-critic-target}--\ref{algo:update-actor-target}). Finally, we update the parameters $\theta^{Q_c}$ of the DDQL Q-network using the Bellman equation in line (\ref{algo:update-ddql-q-network}) after selecting the action using the target Q-network $Q'_c(\bm{s}, a | \theta^{Q'_c})$ in line \ref{algo:ddql-action-selection}. Similar to the DDPG target network, we update in line \ref{algo:update-ddql-target} the DDQL target Q-network every $P$ iterations.
\begin{algorithm}
\small
\caption{Hybrid DDPG-DDQL algorithm for antenna allocation and MIMO precoding}\label{algo:DRL-for-free-cell-network}
\begin{algorithmic}[1]
\State Randomly initialize the critic $Q(\bm{s}, a | \theta^{Q})$ and the actor $\mu(\bm{s} | \theta^{\mu})$ with weights $\theta^{Q}$ and $\theta^{\mu}$ \label{algo:init1}
\State Initialize target network $Q'$ and $\mu'$ with weights $\theta^{Q^{\prime}} \leftarrow \theta^{Q}$, $\theta^{\mu^{\prime}} \leftarrow \theta^{\mu}$ \label{algo:init2}
\State Randomly initialize the $Q_c$-network $Q_c(\bm{s}, a | \theta^{Q_c})$ \label{algo:init3}
\State Initialize the target network $Q'_c(\bm{s}, a | \theta^{Q'_c})$ with weights $\theta^{Q_c^{\prime}} \leftarrow \theta^{Q_c}$ \label{algo:init4}
\State Initialize replay buffer R \label{algo:init5}
\For {$episode=1,\dots, E$}
\State Receive initial observation state $s_1$ after initializing the environment \label{algo:init-env}
\For {$t=1,\dots, T$}
\State  Select the MIMO precoding action $a^{p}_{t}=\mu\left({\bm{s}}_{t} | \theta^{\mu}\right)$ \label{algo:pick-ab}
\State  Select the antennas' allocation action $a^{\mathcal{C}}_t=\arg \max _{a^{\mathcal{C}}} Q_c\left(\bm{s}_t, a^{\mathcal{C}}\right)$ \label{algo:pick-ac}
\State Define $a_t = (a^p_t, a^{\mathcal{C}}_t)$ \label{algo:combine-ab-ac}
\State Execute action $a_{t}$ and observe reward $r_{t}$ and observe new state $\bm{s}_{t+1}$ \label{algo:get-state-reward}
\State Store transition $(\bm{s}_{t}, a_{t}, r_{t}, \bm{s}_{t+1})$ in $R$ \label{algo:store-transition}
\State Sample a random minibatch of $\mathcal{L}$ transitions $(\bm{s}_{i}, a_{i}, r_{i}, s_{i+1})$ from $R$ \label{algo:sample-replay-buffer}
\State Get $a^p_i$ and $a^{\mathcal{C}}_i$ from $a_i$
\Statex\LeftComment{2}{\underline{Training the DDPG networks}}
\State Compute the TD target $y_{i}=r_{i}+\zeta\, Q^{\prime}\left(\bm{s}_{i+1}, \mu^{\prime}\left(\bm{s}_{i+1} | \theta^{\mu^{\prime}}\right) | \theta^{Q^{\prime}}\right)$  \label{algo:ddpg-td-target}
\State Update the critic $Q(\bm{s}, a | \theta^{Q})$ by minimizing the loss: $L=\frac{1}{\mathcal{L}} \sum_{i}\left(y_{i}-Q\left(\bm{s}_{i}, a^p_{i} | \theta^{Q}\right)\right)^{2}$ \label{algo:update-critic}
\Statex \qquad \quad  Update the actor policy $\mu(\bm{s} | \theta^{\mu})$ using a monte-carlo approximation of (\ref{eq:gradient-loss-function-actor}):
\State \qquad \qquad$\left.\left.\nabla_{\theta^{\mu}} J \approx \frac{1}{\mathcal{L}} \sum_{i} \nabla_{a} Q\left(\bm{s}, a | \theta^{Q}\right)\right|_{\mathcal{S}=\bm{s}_{i}, a=\mu\left(\bm{S}_{i}\right)} \nabla_{\theta^{\mu}} \mu\left(\bm{s} | \theta^{\mu}\right)\right|_{\mathcal{S}=\bm{s}_{i}}$
\Statex \qquad \quad Update the DDPG target networks $Q'$ and $\mu'$ \textbf{if} $mod(t, P)=0$:
\State \qquad \qquad ${\theta^{Q^{\prime}} \leftarrow \tau \theta^{Q}+(1-\tau) \theta^{Q^{\prime}}}$ \label{algo:update-critic-target}
\State \qquad \qquad ${\theta^{\mu^{\prime}} \leftarrow \tau \theta^{\mu}+(1-\tau) \theta^{\mu^{\prime}}}$ \label{algo:update-actor-target}
\Statex\LeftComment{2}{\underline{Training the DDQL networks}}
\State select $a^{*}=\arg \max _{a} Q'_c\left(\bm{s}_{i+1}, a| \theta^{Q'_c}\right)$   \label{algo:ddql-action-selection}
\Statex \qquad \quad Update the $Q_c$ usin{}g:
\State \qquad \qquad $Q_c(\bm{s}_i, a^{\mathcal{C}}_i | \theta^{Q_c}) \leftarrow Q_c(\bm{s}_i, a^{\mathcal{C}}_i| \theta^{Q_c})+\nu\,\left(r_i+\zeta\, Q_c\left(\bm{s}_{i+1}, a^{*}| \theta^{Q_c}\right)-Q_c(\bm{s}_i, a^{\mathcal{C}}_i| \theta^{Q_c})\right)$ \label{algo:update-ddql-q-network}
\Statex \qquad \quad Update the DDQL target networks $Q_c'$ \textbf{if} $mod(t, P)=0$:
\State \qquad \qquad ${\theta^{Q_1^{\prime}} \leftarrow \tau \theta^{Q_1}+(1-\tau) \theta^{Q_1^{\prime}}}$ \label{algo:update-ddql-target}
\EndFor

\EndFor
\end{algorithmic}
\end{algorithm}
\section{Performance Results}  

\subsection{Simulation Parameters}
We provide some numerical results to discuss the performance of the proposed scheme under different system parameters.
Each value  is  obtained via  $1\times10^6$ Monte-Carlo simulation runs. 
We assume that the channel fading coefficients in the $\mathcal{P}_1 \leftrightarrow \mathcal{P}_2$ link follow an i.i.d. Rayleigh distribution while the channel fading coefficients for SI follow an i.i.d. Rician distribution.
 
Table \ref{Table3} presents the main network parameters used to obtain the simulation results.
\begin{table}[h!]
    \centering
      \caption{Simulation parameters}
\begin{tabular}{|c|c|}
\hline
Parameter & Value  \\
\hline
\hline
AWGN PSD &
$-169$ dBm/Hz  \\
\hline
$P_S$ & Variable\\
\hline
QoS power, $P_Q$ & $P_S$
\\
\hline
EH and IT Trade-off factor, $\alpha$  & $0.5$\\
\hline
Time Switching factor, $\tau$ & 0.5
\\
\hline
\end{tabular}
    \label{Table3}
\end{table}

\subsection{Results}
\subsubsection{Sub-Optimal Solution}
\begin{figure}[!htb]
		\centering
	\includegraphics[height=8.5 cm, width=9.9 cm]{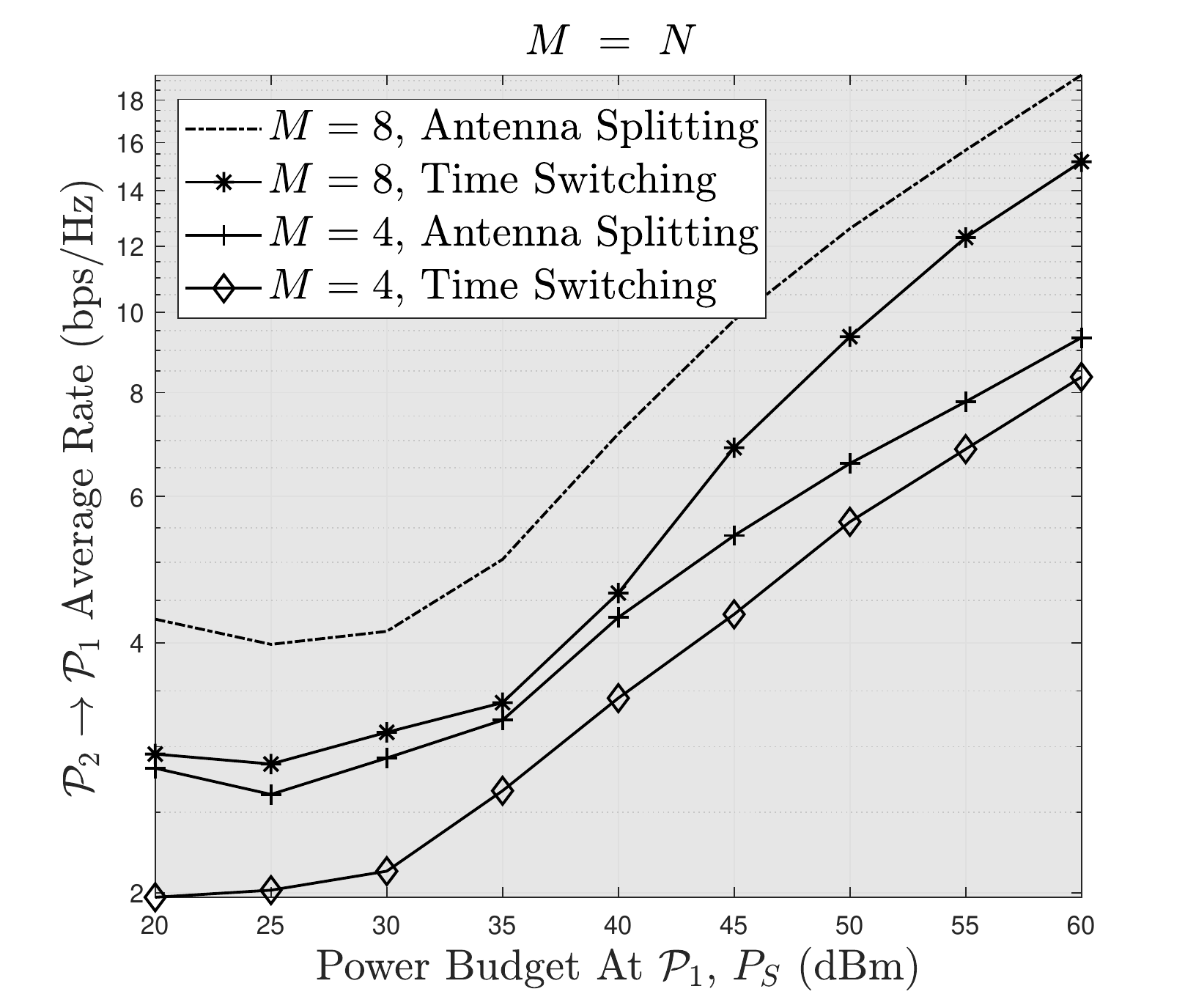}
		\caption{Average rate due to antenna allocation and due to time switching.  }\label{Ant_Spl_Vs_Tim_Sw}
	\end{figure}

This section presents and discusses some insightful results on the performance of the proposed sub-optimal antenna allocation and relaxation-based precoding solution.  
First, in Fig. \ref{Ant_Spl_Vs_Tim_Sw}, we evaluate the performance gain achieved due to the proposed antenna allocation-based EH scheme over the conventional harvest-then-transmit scheme (i.e. time switching-based SWIPT method).  

It can be noticed that the proposed antenna allocation-based EH model achieves a significant rate enhancement in the $\mathcal{P}_1\rightarrow \mathcal{P}_2$ link with largest rate enhancement (around $1.5-2$ bps/Hz) achieved at the low-to-moderate transmission power ranges (within the $20-40$ dBm range).

\begin{figure}[!htb]
		\centering
	\includegraphics[height=8.5 cm, width=9.9 cm]{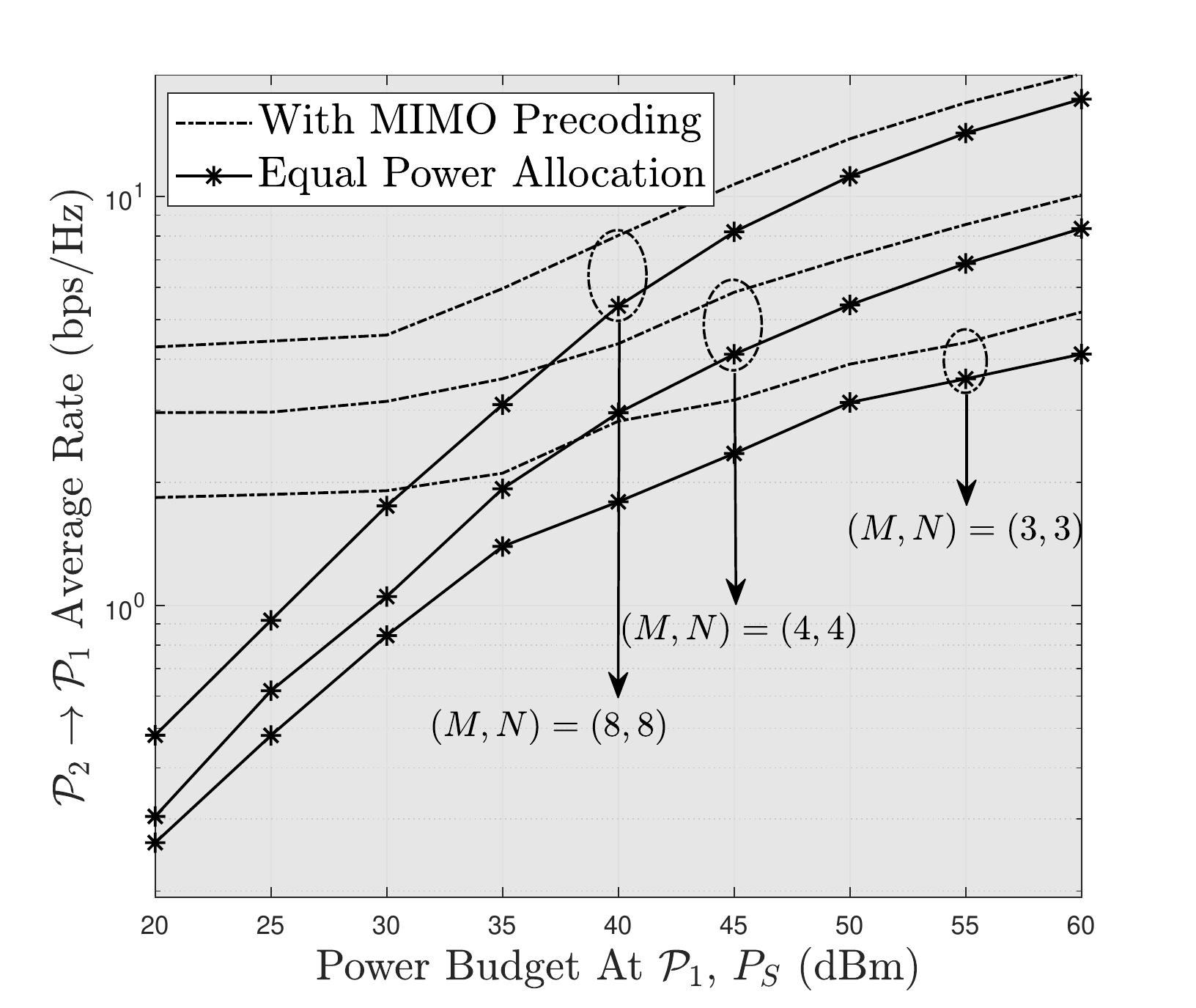}
		\caption{Average rate with MIMO precoding and with equal power allocation.  }\label{Prec_Vs_Eq_Power}
	\end{figure}

Fig. \ref{Prec_Vs_Eq_Power} shows the average transmission rate at the $\mathcal{P}_2\rightarrow \mathcal{P}_1$ link with MIMO-FD precoding and with equal power allocation schemes under different number of antennas.  
As can be noticed from this figure, the best MIMO-FD precoding is achieved at the low-to-moderate transmission power range ($20-40$ dBm). Furthermore, as the maximum power budget at $\mathcal{P}_1$ increases, the gap in performance with the two cases decreases significantly.

\begin{figure}[!htb]
		\centering
	\includegraphics[height=8.5 cm, width=9.9 cm]{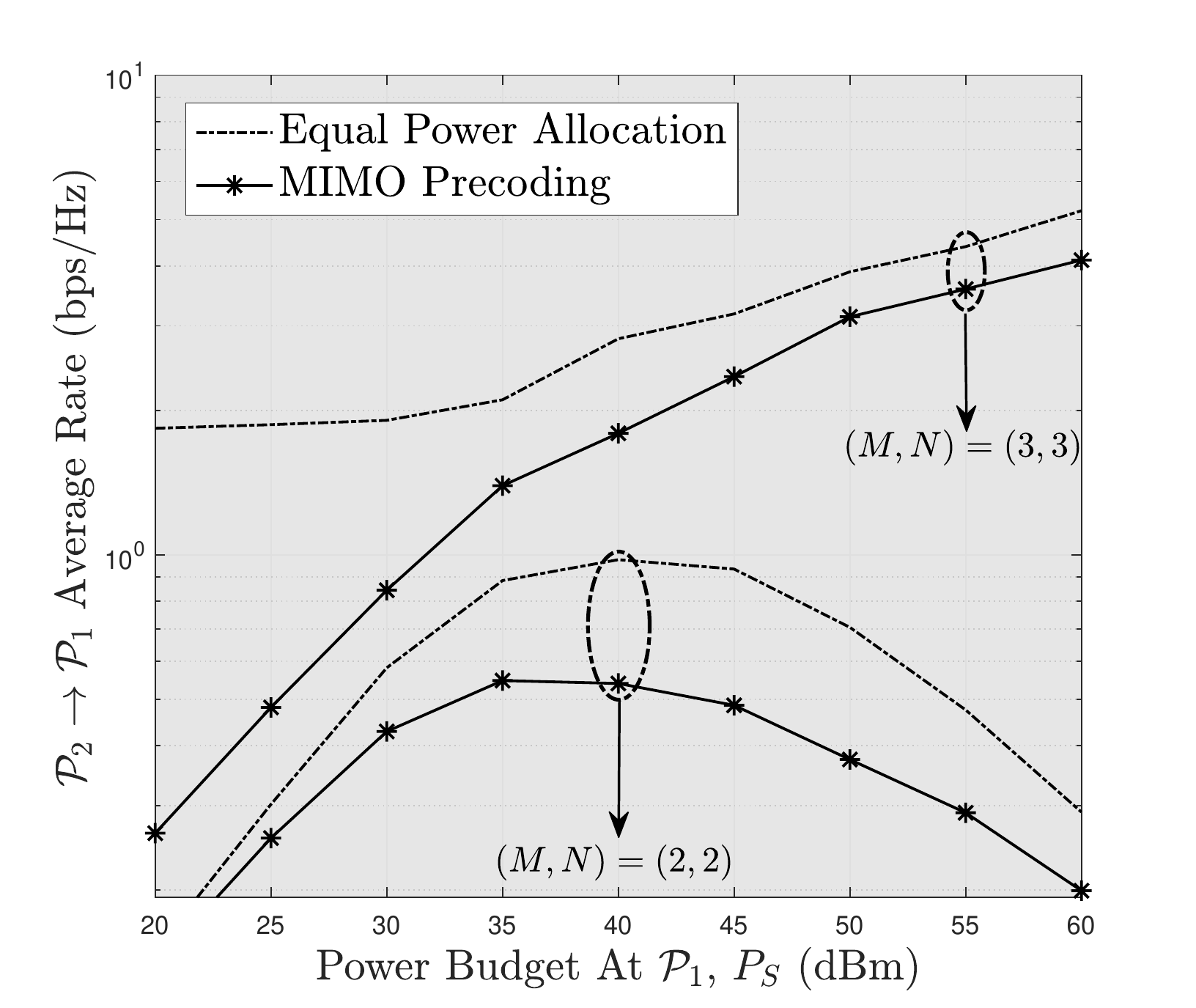}
		\caption{Average rate vs. power budget.  }\label{Ant_Spl_Vs_M}
	\end{figure}

In Fig. \ref{Ant_Spl_Vs_M}, we show the minimum allowable number of antennas ($M$ and $N$) for the MIMO-P2P-FD system for simultaneous EH and IT. 
As can be noticed from this figure, for $(M,N)=(2,2)$, increasing the power budget at $\mathcal{P}_1$ causes the system rate to drop to a small value and a total outage of service occurs at $P_{\text{s}} = 55$ dBm and beyond. This is due to the fact that the $(M,N)=(2,2)$ system has only one possible MIMO subsystem configuration with $(M_I,N_I)=(1,1)$ and $(M_h,N_h)=(1,1)$, and therefore, no precoding scheme can be used to reduce the SI component at $\mathcal{P}_1$.

\subsubsection{DRL-Based Solution}	
\begin{figure}[!htb]
		\centering
	\includegraphics[height=8.5 cm, width=9.9 cm]{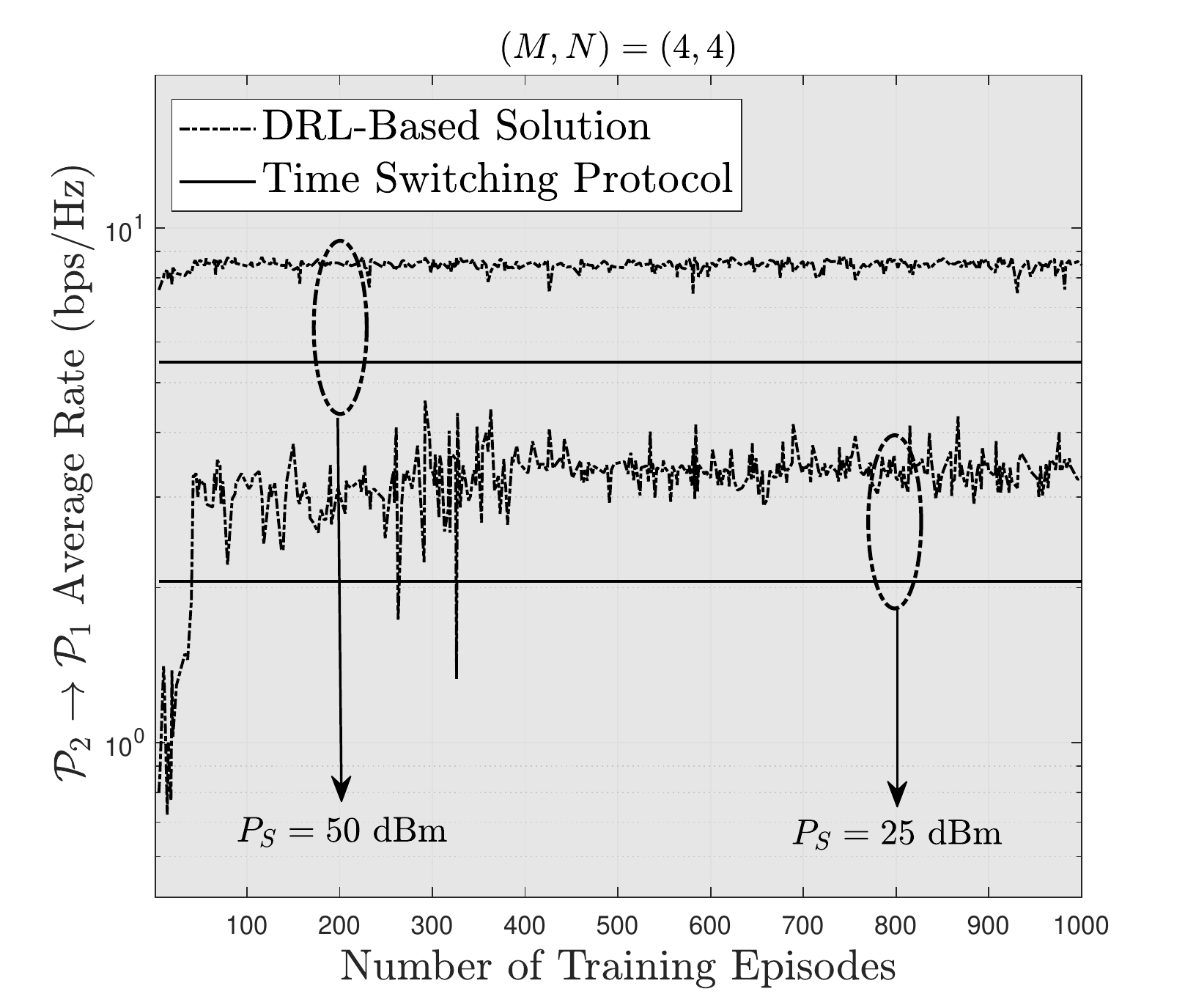}
		\caption{Average transmission rate for DRL-based method and time switching-based SWIPT method.  }\label{DRL_Vs_Tim_Sw}
	\end{figure}

Here, we study the performance of using the proposed DRL model in solving problem $\textbf{P}_1$ in (\ref{GeneralProb}).
First, in Fig. \ref{DRL_Vs_Tim_Sw}, we compare the performance of the proposed antenna allocation-based EH scheme solved by the DRL-based method with that of cthe onventional time switching-based EH scheme solved by the conventional one-way MIMO precoding scheme.
It can be noticed that a significant performance enhancement can be achieved by the proposed antenna allocation-based EH model compared to that of conventional time switching-based model (around $1.5$ bps/Hz at $P_s = 25$ dBm and $3$ bps/Hz at $P_s = 50$ dBm). Note that this performance gain is larger than that achieved by the antenna allocation-based scheme when solved using the proposed sub-optimal antenna allocation with relaxation-based precoding solution (around $1$ bps/Hz at $P_s = 25$ dBm and $1.5$ bps/Hz at $P_s = 50$ dBm).   

\begin{figure}[!htb]
		\centering
	\includegraphics[height=8.5 cm, width=9.9 cm]{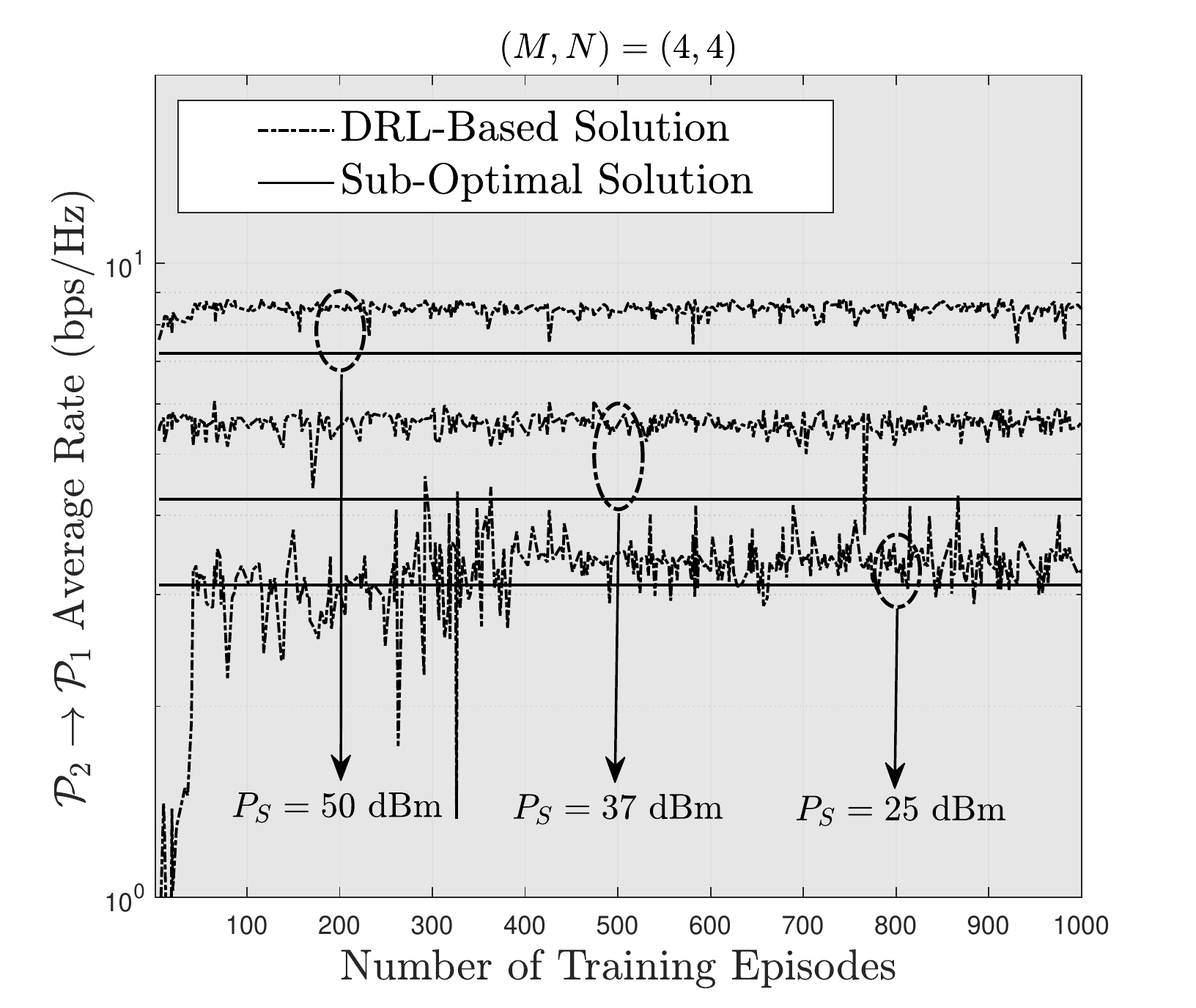}
		\caption{Average rate of DRL-based solution vs. sub-optimal antenna allocation with relaxation-based precoding.  }\label{DRL_Vs_Math}
	\end{figure}

To further emphasize the superiority of the DRL-based solution for the proposed antenna allocation-based EH system, Fig. \ref{DRL_Vs_Math} shows the average normalized transmission rate in the $\mathcal{P}_2 \rightarrow \mathcal{P}_1$ link due to the proposed sub-optimal antenna allocation with relaxation-based precoding method and the DRL-based method. 
It can be noticed from this figure that the DRL-based solution shows a better performance than the sub-optimal antenna allocation with relaxation-based precoding under all power ranges (especially for a medium power value $P_s = 37$ dBm). This is due to the fact that the DRL-based solution learns all possible MIMO subsystem configurations and their corresponding precoding matrices. However, the sub-optimal antenna allocation with relaxation-based precoding solution only approximates the best MIMO subsystem configuration and solves the MIMO-FD precoding problem separately (after deciding on the MIMO subsystem configuration) without trying all possible combinations.
\section{Conclusion}

A MIMO-enabled full-duplex (FD) point-to-point (P2P)  communications scheme with energy harvesting (EH) has been studied. 
Specifically, in this scheme, one device is assumed to transmit an energy signal to a second device that is equipped with an EH unit to harvest the received ES. 
In return, the second device utilizes the harvested energy to transmit an information signal to the first device. 
Both the devices are equipped with multiple antennas with every antenna used either for EH or IT.
After formulating the problem of optimal antenna selection and power allocation for this system, two solution methods have been proposed. The first method is based on sub-optimal antenna splitting and a relaxation-based precoding scheme. The second solution is based on a hybrid deep reinforcement learning-based (DRL)-based implementation, namely, a deep deterministic policy gradient (DDPG)-deep double Q-network (DDQN) model. Performances of both the methods have been studied numerically and the DRL-based method has been observed to provide a superior performance.
 One possible extension of this work is to extend the proposed antenna allocation and precoding model to a multi-user scenario.

\bibliographystyle{IEEEtran}
\bibliography{IEEEabrv,yasser}


\end{document}